\documentclass[aps,twocolumn,superscriptaddress,floatfix]{revtex4-2}
\usepackage{amsmath,amssymb,graphicx,graphics,color}
\usepackage{dcolumn}  
\usepackage{bm}       

\newcommand{\br}{{\bm r}}
\newcommand{\bF}{{\bm F}}
\newcommand{\rd}{\mathrm{d}}

\newcommand{\ra}{\rangle}
\newcommand{\be}{\begin{equation}}
\newcommand{\ee}{\end{equation}}
\newcommand{\bes}{\begin{eqnarray}}
\newcommand{\ees}{\end{eqnarray}}
\newcommand{\Ho}{\hat{H}}
\newcommand{\Uo}{\hat{U}}
\newcommand{\no}{\hat{n}}
\newcommand{\bo}{\hat{b}^{\phantom{\dag}}}
\newcommand{\ba}{\hat{b}^\dag}
\renewcommand{\ao}{\hat{a}^{\phantom{\dag}}}
\renewcommand{\aa}{\hat{a}^\dag}

\begin{document}

\title{The optimal frequency window for Floquet engineering in optical lattices}
\author{Gaoyong Sun}
\email{gysun@nuaa.edu.cn}
\affiliation{Max-Planck-Institut f\"ur Physik komplexer Systeme, N\"othnitzer Stra\ss{}e 38, 01187 Dresden, Germany}
\affiliation{College of Science, Nanjing University of Aeronautics and Astronautics, 211106, China}
\author{Andr\'e Eckardt}
\email{eckardt@pks.mpg.de} 
\affiliation{Max-Planck-Institut f\"ur Physik komplexer Systeme, N\"othnitzer Stra\ss{}e 38, 01187 Dresden, Germany}

\begin{abstract}
The concept of Floquet engineering is to subject a quantum system to 
time-periodic driving in such a way that it acquires interesting novel properties. 
It has been employed, for instance, for the realization of artificial magnetic fluxes in optical lattices and, typically, 
it is based on two approximations. First, the driving frequency is assumed to 
be low enough to suppress resonant excitations to high-lying states above some 
energy gap separating a low energy subspace from excited states.
Second, the driving frequency is still assumed to be large compared to the 
energy scales of the low-energy subspace, so that also resonant excitations 
within this space are negligible.
Eventually, however, deviations from both approximations will lead to unwanted 
heating on a time scale $\tau$. Using the example of a one-dimensional system 
of repulsively interacting bosons in a shaken optical lattice, we investigate 
the optimal frequency (window) that maximizes $\tau$. As a main result, we find 
that, when increasing the lattice depth, $\tau$ increases faster than the 
experimentally relevant time scale given by the tunneling time $\hbar/J$,
so that Floquet heating becomes suppressed.
\end{abstract}


\maketitle


\section{Introduction}
The idea of Floquet engineering is to subject a quantum system to time-periodic
driving in such a way that it acquires interesting novel properties that are 
difficult to achieve by other means. This concept has been applied very 
successfully to systems of atomic quantum gases in optical lattices 
\cite{Eckardt17}. The fact that these systems are extremely clean, well 
isolated from their environment, and highly tunable also in a time-dependent 
fashion makes them an ideal platform for studying coherent many-body dynamics. 
Examples for Floquet engineering in optical lattices include, among others, 
dynamic localization \cite{DunlapKenkre86, LignierEtAl07}, 
photon-assisted tunneling \cite{EckardtHolthaus07, SiasEtAl08, IvanovEtAl08, 
AlbertiEtAl09, HallerEtAl10}, the control of an interaction-induced quantum phase 
transition \cite{EckardtEtAl05b, ZenesiniEtAl09}, the creation of kinetic 
frustration \cite{EckardtEtAl10, StruckEtAl11}, artificial magnetic fields
\cite{Kolovsky11, BermudezEtAl11, AidelsburgerEtAl11, HaukeEtAl12b, StruckEtAl12,
StruckEtAl13, AidelsburgerEtAl13, MiyakeEtAl13, AtalaEtAl14, KennedyEtAl15},  
topological band structures \cite{OkaAoki09, JotzuEtAl14, AidelsburgerEtAl15}, 
and number dependent gauge potentials \cite{GoergEtAl19}.

A simple explanation of the basic concept underlying Floquet engineering is often
given by considering the one-cycle time-evolution operator 
\be
\Uo(T,0) = \mathcal{T}\exp\bigg[\frac{1}{i\hbar}\int_0^T\!\rd t\, \Ho(t)\bigg]
\ee
for a quantum system described by a time-periodic Hamiltonian with angular driving frequency $\omega=2\pi/T$, 
\be
\Ho(t)=\Ho(t+T),
\ee
where $\mathcal{T}$ denotes time ordering. The fact that this operator is unitary
allows one, at least formally, to express it in terms of an hermitian 
operator $\Ho_F$ that is called Floquet Hamiltonian, 
\be
\Uo(T,0)\equiv \exp\Big(\frac{1}{i\hbar}T\Ho_F\Big).
\ee
This effective  time-independent Hamltonian $\Ho_F$ governs the time evolution of
the system, when it is monitored stroboscopically in integer steps of the driving
period $T$. Thus, at a first glance, one might expect that the driven system 
behaves as some effective non-driven system described by the Hamiltonian $\Ho_F$. 
However, while the above reasoning applies to small quantum systems, the situation
in many-body systems is more complex. Here the eigenstates of $\Ho_F$ will 
typically be superpositions of states having very different energies. This is a 
consequence of the lack of energy conservation in driven systems, which is 
reflected in the possibility of resonant coupling, and the fact that in a large 
system resonances will be ubiquitous. The lack of energy conservation suggests 
that in the thermodynamic limit the system approaches an infinite-temperature-like 
state, so that in the sense of eigenstate thermalization the eigenstates of
$\Ho_F$ represent an infinite-temperature ensemble \cite{LazaridesEtAl14b, 
DAlessioRigol14}. From this point of view, the Floquet Hamiltonian $\Ho_F$
does not seem to be a suitable object for engineering interesting system properties.

The fact that Floquet engineering can, nevertheless, be a useful concept also
in many-body quantum systems, is related to the observation that in some
parameter regimes the time scale $\tau$ associated with unwanted resonant 
processes, where the system absorbs (or emits) energy quanta $\hbar\omega$, can 
become rather long. Since typically energy absorption dominates, in the 
following, we will denote such detrimental energy-non-conserving processes as
``heating'' and $\tau$ as the corresponding ``heating'' time \footnote{This 
(rather common) terminology shall not imply that the system is described by 
thermodynamic variables such as temperature.}  
On times shorter than $\tau$, we might be able to engineer and study interesting 
driving-induced physics described by an approximate time-independent effective 
Hamiltonian $\Ho_\text{eff}$, corresponding to a non-driven system with modified 
properties. The standard strategy employed for deriving such an effective 
Hamiltonian involves two steps \cite{EckardtEtAl05b}:

The first step is given by a \emph{low-frequency approximation}, where the 
assumption is made that the system remains in a low-energy subspace, which is 
separated by an energy gap from excited states, which is much larger than the 
driving frequency. In non-driven systems, such low-energy approximations are 
common. For example, in a lattice system higher-lying orbital states spanning 
Bloch bands above a band gap are neglected, when deriving Hubbard-type 
tight-binding models, or doublon-holon excitations lying above a charge gap of 
a Mott insulator are eliminated adiabatically, in order to derive spin 
Hamiltonians. For a sufficiently large energy gap, in non-driven systems one 
can expect that the admixture of higher-lying states is captured by a 
converging perturbation theory and will always remain small. In contrast, in a 
periodically driven many-body system, the situation is generically different. 
Here resonant excitations to the neglected excited states can occur, where the 
drive provides one or several energy quanta $\hbar\omega$. Such processes 
contribute to the aforementioned detrimental heating. However, for driving 
frequencies (and amplitudes) much lower than the gap, so that the system would 
need to absorbe many energy quanta $\hbar\omega$ (``photons'') at once, they 
can be exponentially slow with respect to the photon number. Thus, by 
estimating the associated heating rate \cite{ChoudhuryMueller14, WeinbergEtAl15,
ChoudhuryMueller15, GenskeRosch15, BilitewskiCooper15, BilitewskiCooper15b, 
StraeterEckardt16, ReitterEtAl17, RajapopalEtAl19, SinghEtAl19}, we might be 
able to argue that we can still neglect higher-lying states on the time scale of 
an experiment.

The second step is given by a \emph{high-frequency approximation}. Let us 
assume that according to the first step we are able to neglect, say, higher-
lying Bloch bands, so that we can describe our system by a Hubbard Hamiltonian 
acting in the lowest Bloch band. Now, the periodic drive can still resonantly 
create excitations within this low-energy subspace. This form of energy 
absoprtion (heating) can be reduced considerably by considering driving 
frequencies that are sufficiently large, so that absorbing an energy quantum of 
$\hbar\omega$ corresponds to an exponentially slow high-order process in which 
several elementary excitations are created at once \cite{EckardtHolthaus08b, 
EckardtAnisimovas15}. If this is the case, we can employ a rotating-wave 
approximate and describe the system by the time-averaged low-energy Hamiltonian 
(or compute also further corrections using a high-frequency expansion 
\cite{CasasEtAl00, VerdenyEtAl13, GoldmanDalibard14, EckardtAnisimovas15, 
BukovEtAl15}). In this way, we arrive at an approximate effective Hamiltonian 
$\Ho_\text{eff}$ that describes the dynamics of our system on time scales 
before driving-induced heating sets in. The leading order of this expansion is 
simply given by the time-averaged Hamiltonian and corresponds to a 
rotating-wave approximation.

The two steps outlined above require that there is a window of suitable driving
frequencies that are both low compared to the relevant energy gap separating the 
low-energy subspace from higher-lying states and large compared to the energy
scales governing this low-energy subspace. In this article, we investigate the
question, whether such an optimal frequency window exists, using the 
experimentally relevant example of repulsively interacting bosonic atoms in a 
periodically shaken one-dimensional optical lattice. For this purpose, we 
compare the evolution generated by an approximate effective Hamiltonian 
$\Ho_\text{eff}$ acting in the lowest Bloch band to the evolution obtained from 
integrating the dynamics of the fully time-dependent model that, apart from the 
lowest band, contains also first excited band. 

The remaining part of this paper is organized as follows: After introducing 
the system and the model in Sec.~\ref{sec:system}, in Sec.~\ref{sec:Heff} we 
recapitulate the derivation of the approximate effective Hamiltonian
$\Ho_\text{eff}$ from the low and the high-frequency approximation. In the 
following two sections, we then compare the evolution generated by 
$\Ho_\text{eff}$ to numerical simulations: in Sec.~\ref{sec:intra} we 
investigate the break-down of the high-frequency approximation due to intraband 
heating and in Sec.~\ref{sec:intrainter} we study the combined effect of 
intraband and interband heating beyond the high and low frequency 
approximation. Finally, we close with Sec.~\ref{sec:conclusions}.

\section{System and Model\label{sec:system}}
We consider a system of ultracold bosonic atoms in a one-dimensional optical 
lattice potential
\be\label{fig:lattice}
V(\br) = V_0\sin^2(k_L x)+V_\perp(y,z).
\ee
Here the laser wave number $k_L$ defines the recoil energy 
$E_R=\hbar^2k_L^2/(2m)$ with atom mass $m$, corresponding to the kinetic energy 
required to localize a particle on the length of a lattice constant $a=\pi/k_L$. 
Typical recoil energies take values of a few kHz. The deep confining potential 
$V_\perp(y,z)\simeq\frac{m}{2}\omega_\perp^2(y^2+z^2)$ shall reduce the 
dynamics to one spatial dimension via a large transverse excitation gap
$\hbar\omega_\perp$ that freezes the particles in the lowest transverse
single-particle state. More precisely, $\omega_\perp$ will be chosen large 
enough, so that the time scale for driving induced transverse heating can be 
expected to be much longer than the one for resonant excitations of 
longitudinal degrees of freedom in lattice direction, which we are going to 
investigate here. 

The system shall be driven periodically in time by the homogenous sinusoidal 
force pointing in the lattice direction ${\bm e}_x$,
\be\label{fig:force}
\bF(t) = - K a\cos(\omega t) {\bm e}_x.
\ee
It is characterized by the driving strength $K$, corresponding to the amplitude 
of the potential offset between neighboring lattice sites, and the angular 
driving frequency $\omega$, which defines also the driving period $T=2\pi/\omega$.
Such a force can be realized as an inertial force by shaking the lattice back and 
forth in $x$ direction. 

In the absence of periodic forcing, experiments performed in the regime of deep 
lattices, $V_0/E_R\gtrsim 5$, at the typical ultracold quantum gas temperatures
are described accurately by the single-band Bose Hubbard model \cite{JakschEtAl98}
\be\label{eq:Hs}
\Ho_s = -J_s\sum_{\ell=1}^{M-1}
             \Big(\ba_{s\ell+1} \bo_{s\ell} + \text{H.c.} \Big)
                +\frac{U_s}{2} \sum_{\ell=1}^{M}\no_{s\ell}(\no_{s\ell}-1)  .
\ee
Here the index $\ell$ denotes the lattice sites in ascending order from 1 to $M$ and the label $s$ indicates the lowest Bloch band to be distinguished from the first excited band, labeled by $p$, which is considered below. Moreover,
$\ba_{\alpha\ell}$, $\bo_{\alpha \ell}$, and 
$\no_{\alpha\ell}=\ba_{\alpha\ell}\bo_{\alpha\ell}$ denote the creation, 
annihilation and number operator for a boson in a Wannier state of band $\alpha$ 
on site $\ell$. Nearest-neighbor tunneling is described by the parameter $J_s$ 
and on-site interactions by the Hubbard parameter $U_s$.

While in a non-driven system, a description in the low-energy subspace of the $s$
band is well justified, this assumption is not as clear in a system that is 
driven periodically. Even if the driving frequency is small compared to the 
band gap separating the $s$ band from the first excited $p$ band, states of 
excited bands might still be populated via multiphoton excitations corresponding 
to either single-particle processes \cite{WeinbergEtAl15, StraeterEckardt16} or 
two-particle scattering \cite{ReitterEtAl17}. If periodic driving is used to 
control the physics of the lowest band, such excitation processes must be viewed 
as unwanted heating. In order to estimate this effect, we will also take into 
account the first excited band, which for the undriven lattice is captured 
by the Hamiltonian 
\bes\label{eq:Hp}
\Ho_p &=& \Delta\sum_{\ell=1}^{M}  \no_{p\ell} 
			+ J_p\sum_{\ell=1}^{M-1}\Big(\ba_{p\ell+1} \bo_{p\ell} + \text{H.c.} \Big)
\\\nonumber&&
               +\,\frac{U_p}{2}\sum_{\ell=1}^{M} \no_{p\ell}(\no_{p\ell}-1)  ,
\ees
and coupled to the $s$ band via the interband interaction term
\be\label{eq:Hsp}
\Ho_{sp} =U_{sp}\sum_{\ell=1}^M \Big[2\no_{s\ell}\no_{p\ell} 
                            + \frac{1}{2}\Big(\ba_{p\ell}\ba_{p\ell}
                               \bo_{s\ell}\bo_{s\ell} + \text{H.c.} \Big)\Big].
\ee
Here $\Delta$ denotes the orbital energy required to excite a particle to a 
Wannier state of the $p$ band and $J_p$ and $U_p$ describe nearest-neighbor 
tunneling and on-site interactions in this $p$ band, respectively. The on-site 
scattering and repulsion between $s$ and $p$ states is quantified by $U_{sp}$.

If the energy scales of the periodic force, $\hbar\omega$ and $K$, remain below 
the band gap $\Delta$, the bands of the undriven problem, $s$ and $p$, provide a 
useful basis also for the description of the driven system (see supplemental 
material of Ref.~\cite{ReitterEtAl17}). Assuming this regime, we project the 
potential $-\br\cdot\bF(t)$ induced by the force to the lowest two bands and 
obtain the driving term of the Hamiltonian:
\be\label{eq:Hdr}
\Ho_\text{dr}(t) = K\cos(\omega t) \sum_{\ell=1}^{M} 
                            \Big[\ell \Big(\no_{s\ell}+\no_{p\ell}\Big) 
               +\,\eta\Big( \ba_{p\ell}\bo_{s\ell} + \text{H.c.}\Big)   \Big]
\ee
where $\eta$ is the dipole matrix element between two Wannier states of the $s$ 
and the $p$ band on the same lattice site in units of the lattice constant. 

\begin{figure}[t]
\includegraphics[width=8.6cm]{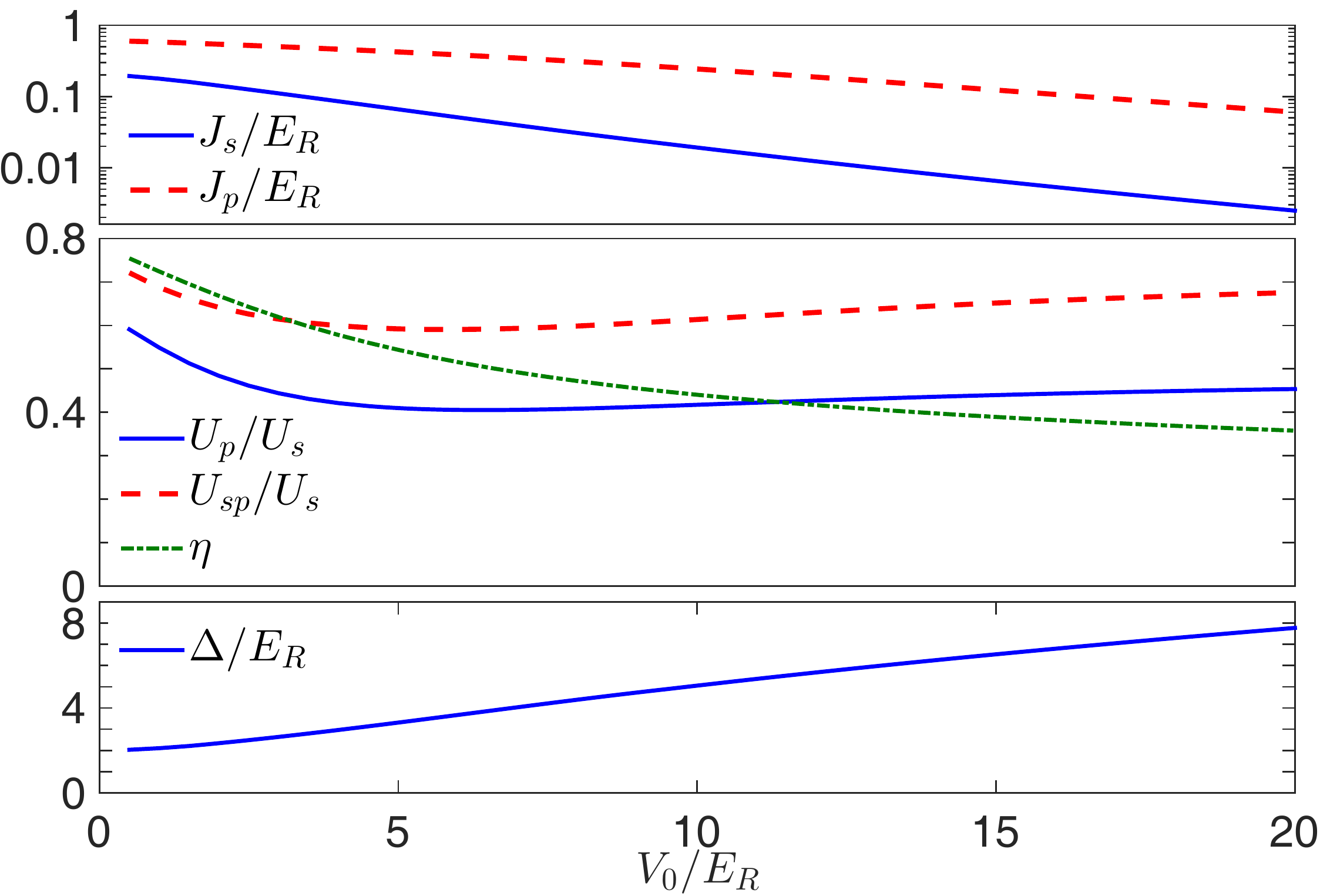}
\caption{Parameters characterizing the two-band Bose-Hubbard model for a shaken 
one-dimensional optical cosine lattice plotted versus the lattice depth
$V_{0}/E_{R}$.}
\label{fig:parameters}
\end{figure}

The total Hamiltonian to be used for our analysis is now given by
\be\label{eq:tb}
\Ho(t) = \Ho_s + \Ho_{p} + \Ho_{sp} + \Ho_\text{dr}(t). 
\ee
The number of independent parameters that describe this model is reduced 
considerably by noticing that $J_s/E_R$, $J_p/E_R$, $\Delta/E_R$, and $\eta$ 
are determined completely by the dimensionless lattice depth $V_0/E_R$. 
Moreover, the interaction parameters $U_s$, $U_p$, and $U_{sp}$ share the very 
same (linear) dependence on both the $s$-wave scattering length $a_s$ (which 
can be tuned using Feshbach resonances) and the transverse confinement
$\omega_\perp$, so that their ratios $U_p/U_s$ and $U_{sp}/U_s$ equally depend 
on $V_0/E_R$ only. Thus, taking $J_s$ and $\hbar/J_s$ as the units for 
energy and time, respectively, the undriven model is characterized by $V_0/E_R$ 
and $U_s/J_s$ as well as by the average number of particles per site $N/M$. The 
periodic driving is furthermore characterized by the dimensionless diving 
strength $K/J_s$ and angular frequency $\hbar\omega/J_s$. The dependence of the 
model parameters on the lattice depth $V_0/E_R$, obtained from band-structure 
calculations, is shown in Fig.~\ref{fig:parameters}.

\section{The low and the high-frequency approximation\label{sec:Heff}}
Most schemes of Floquet engineering in optical lattices 
(such as for example the control of the bosonic Mott transition 
\cite{EckardtEtAl05b, ZenesiniEtAl09}, the implementation of kinetic 
frustration \cite{EckardtEtAl10, StruckEtAl11}, the creation of artificial 
gauge fields \cite{StruckEtAl12, HaukeEtAl12b, Kolovsky11, AidelsburgerEtAl11, 
StruckEtAl13, AidelsburgerEtAl13, KennedyEtAl15}, and the realization of 
Floquet topological insulators \cite{OkaAoki09,JotzuEtAl14,AidelsburgerEtAl15}) 
are based on two approximations: a low-frequency approximation with respect to orbital degrees of 
freedom and a high-frequency approximation with respect to processes occurring in
the lowest band described by $H_s$. 

The low-frequency single-band approximation is based on the assumption that the 
driving frequency and amplitude remain low enough to ensure that the system 
remains in the subspace spanned by the lowest ($s$-type) Wannier-like orbital at 
each lattice site. It roughly requires driving frequencies
\be\label{eq:lf}
\hbar\omega\ll \Delta
\ee
and driving amplitudes $K$ smaller than a threshold value $K_\text{th}$ 
below which multiphoton transitions are expected to be suppressed exponentially 
with the photon number $\Delta/\hbar\omega$ \cite{StraeterEckardt16}. It leads to a 
description of the system in terms of a tight-binding model with a single orbital 
state per lattice site, which in our case is given by the single-band model 
\be\label{eq:Hsb}
\Ho_\text{sb}(t) = \Ho_s + K\cos(\omega t)\sum_{\ell=1}^M \ell \no_{s\ell}. 
\ee

The high-frequency approximation is is based on the assumption that the 
driving frequency is still large compared to the energy scales $J_s$ and $U_s$ 
governing the low-energy model (\ref{eq:Hsb}), 
\be\label{eq:hf}
\hbar\omega \gg J_s, U_s .
\ee
Under these conditions the resonant creation of collective excitations of 
energy $\hbar\omega$ becomes a slow high-order process that can be ngelected on 
sufficiently short time scales. This allow us to describe the system using an 
approximate effective time-independent Hamiltonian obtained from a 
high-frequency expansion \cite{EckardtEtAl05b,
BukovEtAl15, GoldmanDalibard14, EckardtAnisimovas15}. For that purpose, we 
first perform a gauge transformation with the time-periodic unitary operator
\be
\Uo(t) = \exp\bigg(-i\sum_{\ell=1}^M \theta(t)\ell\no_{s\ell}\bigg)
\ee
with $\theta(t) = K/(\hbar\omega)\sin(\omega t)\ell$, which integrates out the 
driving term. The transformed Hamiltonian
$\Ho' = \Uo^\dag\Ho_\text{sb}\Uo -i\Uo^\dag\dot{\Uo}$ reads 
\bes\label{eq:Hprime}
\Ho'(t) &=&
    -J_s\sum_{\ell=1}^{M-1}
       \Big(e^{i\theta(t)}\ba_{s\ell+1} \bo_{s\ell} + \text{H.c.} \Big)
\nonumber\\&&
                +\,\frac{U_s}{2}\sum_{\ell=1}^{M}\no_{s\ell}(\no_{s\ell}-1) .
\ees
The fact that it possesses typical matrix elements that are are small compared to 
$\hbar\omega$ even for large $K\sim\hbar\omega$ justifies the
high-frequency approximation also for strong driving. Its leading order is given
by the rotating-wave approximation, where the system is described by 
the time-averaged Hamiltonian
\bes\label{eq:Heff}
\Ho_\text{eff}&=&\frac{1}{T} \int_0^T\!\rd t \, \Ho'(t) \rd t
\\\nonumber
    &=&
 -J_s^\text{eff}\sum_{\ell=1}^{M-1}
       \Big(\ba_{s\ell+1} \bo_{s\ell} + \text{H.c.} \Big)
      +\frac{U_s}{2}\sum_{\ell=1}^{M}\no_{s\ell}(\no_{s\ell}-1) ].
\ees
Here the effective tunneling matrix element 
\be\label{eq:Jeff}
J_s^\text{eff} =  J_s \mathcal{J}_0(K/\hbar\omega)
\ee
acquired a dependence on the scaled driving amplitude $K/(\hbar\omega)$ described by a Bessel function $\mathcal{J}_n$.
In this way the time evolution of the system's state $|\psi(t)\ra$ is 
approximately described by 
\be
|\psi(t)\ra \approx \Uo(t)e^{-\frac{i}{\hbar}(t-t_0)\Ho_\text{eff}}
                \Uo^\dag(t_0)|\psi(t_0)\ra.
\ee
In particular, we expect
\be\label{eq:psi_n}
|\psi(n T)\ra \approx e^{-\frac{i}{\hbar}nT\Ho_\text{eff}}
               |\psi(0)\ra \equiv |\psi_n^\text{eff}\ra
\ee
for integers $n$, when monitoring the dynamics stroboscopically in steps of the 
driving period at those times $t=nT$, for which $\Uo(nT)=1$. Higher orders of the 
high-frequency expansion will provide relative corrections of the order of
$J_s/\hbar\omega$ to the evolution governed by $\Ho_\text{eff}$
\cite{GoldmanDalibard14, EckardtAnisimovas15}.

The single-band high-frequency approximation, leading to a description of the
system's dynamics in terms of the approximate effective Hamiltonian~(\ref{eq:Heff}),
requires that there is a window of driving frequencies for which both conditions
(\ref{eq:lf}) and (\ref{eq:hf}) are fulfilled. Since with increasing lattice 
depth $V_0/E_R$ both $J_s$ decreases rapidly and $\Delta$ increases moderately
(see Fig.~\ref{fig:parameters}), while the interaction parameter $U_s$ can be 
made small by tuning the $s$-wave scattering length using a Feshback resonance, 
such a window will open for sufficiently large $V_0/E_R$. However, even within
such a frequency window heating will not be suppressed completely and eventually
make itself felt on some time scale $\tau$. This heating time $\tau$ has to be
compared to the typical duration of an experiment, which will be given by some
fixed multiple of the tunneling time $\hbar/J_s$, which in turn increases
exponentially with the lattice depth [asymptotically for deep lattices
$\ln(J_s/E_\text{R})\simeq -2\sqrt{V_0/E_\text{R}}$ \cite{Zwerger03}, see also
Fig.~\ref{fig:parameters}]. Thus, in order to take into account also this
latter effect, in the following we will investigate the behavior of the
dimensionless heating time $\tau J_s/\hbar$. In doing so, we have to keep in
mind that there will also be background heating (resulting from noise, 
three-body collisions, or scattering with background particles), 
which is independent of the periodic driving and happens on some time scale 
$\tau_0$. Assuming $\tau_0\sim 1\text{s}$ ($\sim 10\text{s}$), requiring $\tau_0\gg\hbar/J_s$, and
noting that $E_R \sim  2\pi\cdot\hbar\,3\text{kHz}$ for typical experiments, we
can see from Fig.~\ref{fig:parameters} that the lattice depth is limited to
values $V_0/E_R\lesssim 15$ $(20)$.

\section{Intraband heating\label{sec:intra}}
Let us first investigate the validity of the high-frequency approximation, 
before considering also heating due to the coupling to the first excited band. 
For this purpose we consider the following quench scenario. We assume that the system 
is prepared in the ground state of the undriven Hamiltonian (\ref{eq:Hs}), when 
at time $t=0$ the driving amplitude is switched on abruptly to a finite value $K$.
We integrate the time evolution of the system described by the time-dependent 
single-band Hamiltonian $\Ho_\text{sb}(t)$ [Eq.~(\ref{eq:Hsb})] and compare it to 
the approximate solution $|\psi_n^\text{eff}\ra$ [Eq.~(\ref{eq:psi_n})] obtained 
from the time-averaged single-band Hamiltonian $\Ho_\text{eff}$. For that purpose 
we consider a small system of $N=6$ particles on $M=10$ lattice sites, for which 
we can integrate the time evolution exactly. 

In order to monitor the deviation between the exact time evolution and the 
dynamics predicted by the rotating wave approximation, we consider the 
expectation value
\be\label{eq:n0}
n_{0}(t)=\langle\aa_{s0}\ao_{s0}\rangle
\quad\text{with}\quad
\ao_{s0}=\frac{1}{\sqrt{M}}\sum_{\ell=1}^M \bo_{s\ell},
\ee
which corresponds to the mean occupation of the single-particle state with
quasimomentum $0$ in the $s$ band. 
The difference
\be
\Delta n_0(t) = n_{0}(t)-n^\text{eff}_{0}(t)
\ee
between the exact expectation value and the one obtained within the rotating-wave 
approximation taken at times $t=nT$ with integer $n$ will serve us as an 
indicator for the validity of the approximations made.
While for the results presented in this section, $n_0(t)$ refers to the dynamics 
generated by the time-dependent single-band Hamiltonian~(\ref{eq:Hsb}), later 
on in the following section, $n_0(t)$ will correspond to the dynamics of the 
full driven two-band model (\ref{eq:tb}).

\begin{figure}[t]
\includegraphics[width=8.8cm]{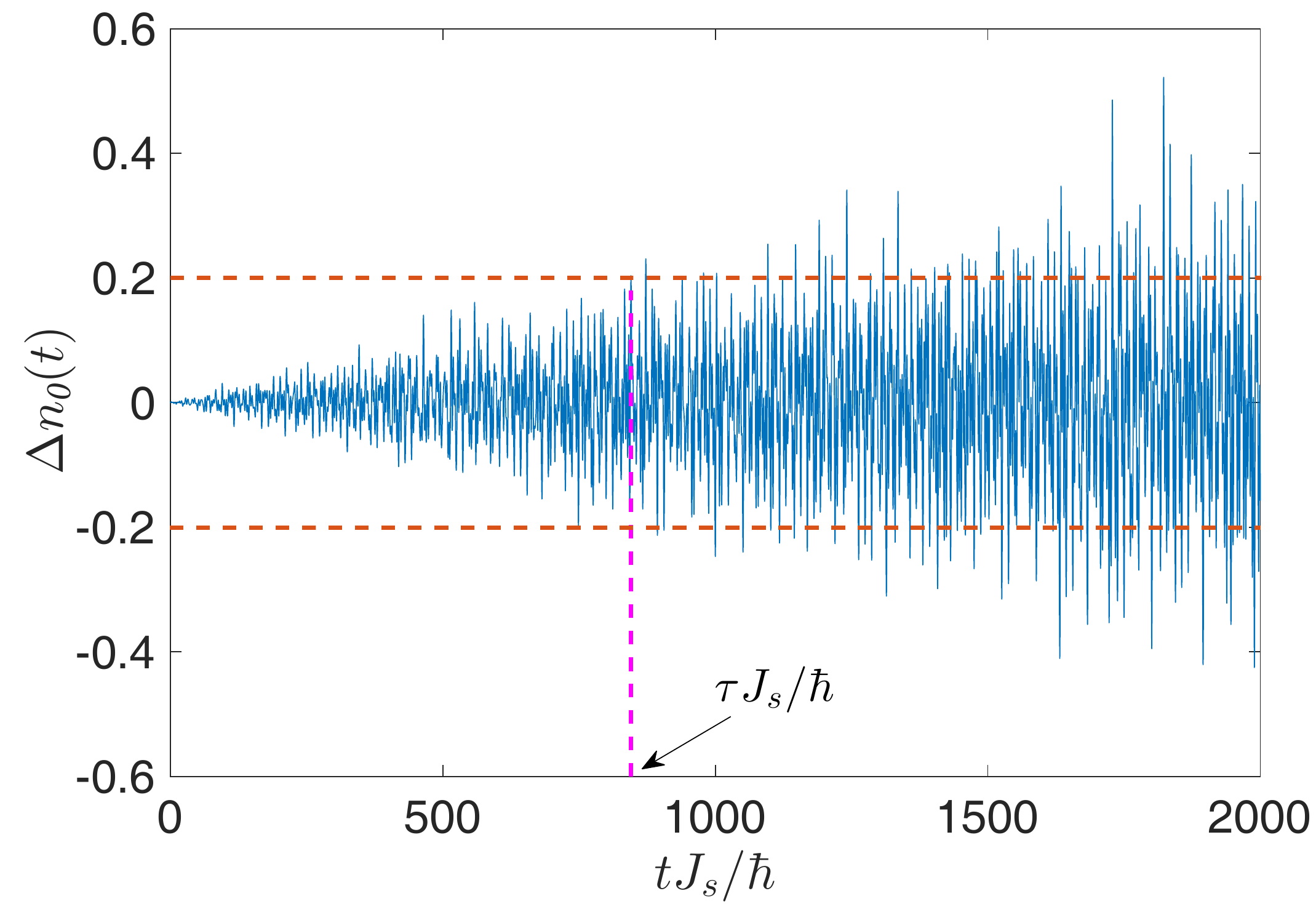}
\caption{Difference $\Delta n_0$ between the exact time
evolution of the single-band Hamiltonian (\ref{eq:Hsb}) and that obtained from
the rotating wave approximation (\ref{eq:psi_n}), taken at times $t=nT$ with inter
$n$. The time evolution is initiated by abruptly switching the amplitude of the drive 
at $t=0$ from $0$ to $K$. The parameters are $N=6$, $M=10$, $V_0/E_R=14$,
$U/J_s=1$, $\hbar\omega/J_s=30$, and $K/\hbar\omega = 4$.
At time $\tau$ the difference $\Delta n_0$ exceeds $0.2$ for the first time.}
\label{fig:evolution}
\end{figure}

In Fig.~\ref{fig:evolution} we plot $\Delta n_0(t)$ for a quench to a 
large driving amplitude $K/\hbar\omega = 4$ (the other parameters are specified 
in the caption). For this value the effective tunneling parameter changes its 
sign, $J_s^\text{eff}\approx -0.4 J$, so that the quench is significant also on 
the level of the rotating wave approximation. We can see that $\Delta n_0(t)$ 
shows an irregular oscillatory behavior, with a roughly linearly growing envelope. 
We define the heating time $\tau$ as the time at which $|\Delta n_0(t)|$ exceeds 
the value $\Delta n_\text{cut}=0.2$ for the first time. Note that $\tau$ gives 
only an estimate for the time scale on which heating starts to play a role. 
The value of $\Delta n_\text{cut}$ is obviously somewhat arbitrary. It is chosen 
to be much smaller than the the initial occupation of the zero momentum state, 
which is of the order of $N$, and it is also smaller than (and of the order of) 
the filling factor $N/M=0.6$ corresponding to the mean occupation of each momentum 
state. The linear spreading of the envelope of $\Delta n_0(t)$ implies that 
altering $\Delta n_\text{cut}$ by a factor of order one will simply alter the 
heating time $\tau$ by roughly the same factor. Note also that the typical 
deviations $|\Delta n_0(t)|$ at time $t=\tau$ are smaller than 
$\Delta n_\text{cut}=0.2$, since in most cases $\Delta n_\text{cut}$ is reached 
the first time during the time evolution when an extreme fluctuation of 
$|\Delta n_0(t)|$ occurs. 

Note that, alternatively, the time $\tau$ could also be defined via the 
(stroboscopic or period-averaged) energy absorption. Such a definition would 
possess the advantage that, to some extent, in experiments it can be measured 
(or at least estimated) directly from time-of-flight images
\cite{WeinbergEtAl15, ReitterEtAl17, RajapopalEtAl19, 
SinghEtAl19}. On the other hand, from the point of view of Floquet engineering, 
the relevant quantity to look at is the deviation from the approximate effective 
Hamiltonian, the physics of which we wish to implement. And these deviations are 
not necessarily proportional to the absorbed energy. Namely, the excitation of a 
particle within the lowest band might be as detrimental as its excitation to the 
first excited band via a multi-photon process, despite the fact that the former 
is associated with a much lower energy absorption than the latter. Therefore, we 
have decided to define the ``heating'' time $\tau$ via deviations from the 
dynamics expected from the target Hamiltonian, as described in the previous 
paragraph.

\begin{figure}[t]
\includegraphics[width=8.9cm]{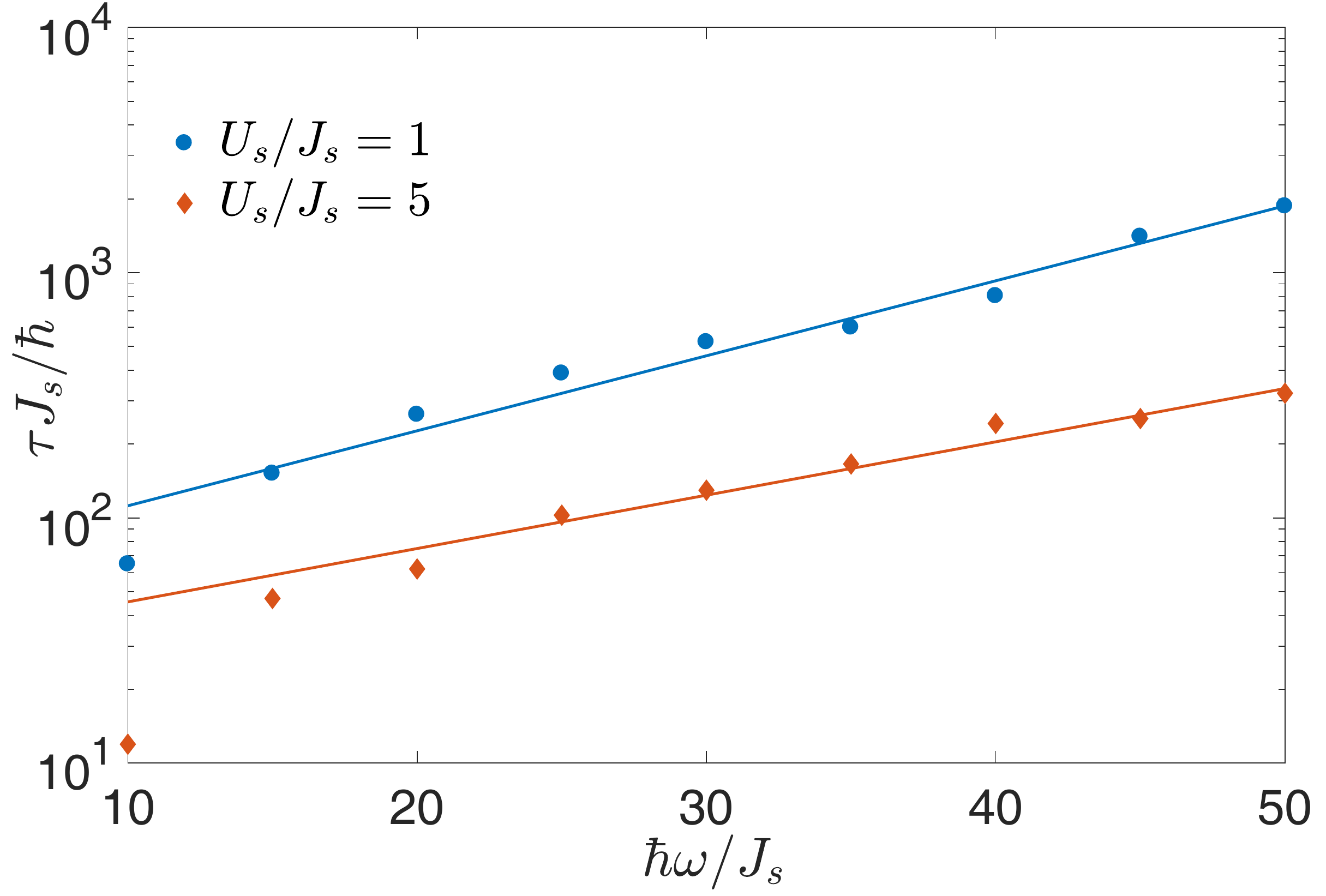}
\caption{Heating time $\tau$ (dots) versus driving frequency $\hbar\omega/J_s$ 
for two different values of the interaction strength $U_s/J_s$. The other 
parameters are chose as in Fig.~\ref{fig:evolution}: $N=6$, $M=10$, 
$K/\hbar \omega=4$, and $V_{R}/E_{R}=14$. The solid lines are exponential fits.}
\label{fig:IntraOmega}
\end{figure}

In Fig.~\ref{fig:IntraOmega}, we plot the heating time $\tau J_s/\hbar$ versus the
driving frequency $\hbar\omega/J$ for two different values of the interaction 
strength $U/J_s=1$ and $U/J_s=5$ (the other parameters are specified in the 
caption). We see that the heating time is considerably reduced for the larger 
value of the interactions. Moreover, an exponential dependence of the heating 
time on the driving frequency can be observed. This agrees with the expectation 
for heating processes based on perturbation theory in Floquet space \cite{
EckardtHolthaus08b}. Namely, one can argue 
that the order of the process of absorbing an energy quantum $\hbar\omega$, 
corresponding to the number of elemantary excitations (quasiparticles) that 
have to be collectively excited, will grow like a power of $\omega$ and that 
the corresponding matrix element will be suppressed exponentially with the 
order \cite{EckardtHolthaus08b, EckardtAnisimovas15}. Such an exponential 
suppression of heating with respect to the driving frequency has recently also 
been proven for spin systems having a finite local energy bound 
\cite{AbaninEtAl15, KuwaharaEtAl16}. Note, however, these proofs do not apply 
to the bosonic Hubbard model considered here, which in principle allows for 
macroscopic site occupations.

\section{Intraband and interband heating\label{sec:intrainter}}
The exponential increase of the heating time with respect to the driving frequency
visible in Fig.~\ref{fig:IntraOmega} is an artifact of the single-band 
description of the driven lattice system. Namely, for sufficiently large driving
frequencies unwanted excitations to higher-lying orbital states (spanning
excited Bloch bands) will become the dominant heating effect. In order to take 
into account this effect, we will include also the coupling to the $p$-band. For 
this purpose we consider the two-band Hamiltonian~(\ref{eq:tb}) and monitor the
heating time $\tau$ defined in the same way as in the previous section. In an
experiment, of course, also further (higher-lying) bands will play a role. 
However, the coupling to the first excited band is most dominant, because with 
respect to the lowest band it both is energetically closest and possesses the 
largest coupling matrix elements. Therefore, the characteristic time scale for 
interband heating processes is determined by transitions to the $p$ band. 
Higher-lying bands can still make themselves felt, e.g.\ in the precise shape 
of resonance lines (as discussed in Ref.~\cite{WeinbergEtAl15}). However, such 
details are not crucial for the present analysis, which is interested in the 
time scales only.

\begin{figure}[t]
\includegraphics[width=8.6cm]{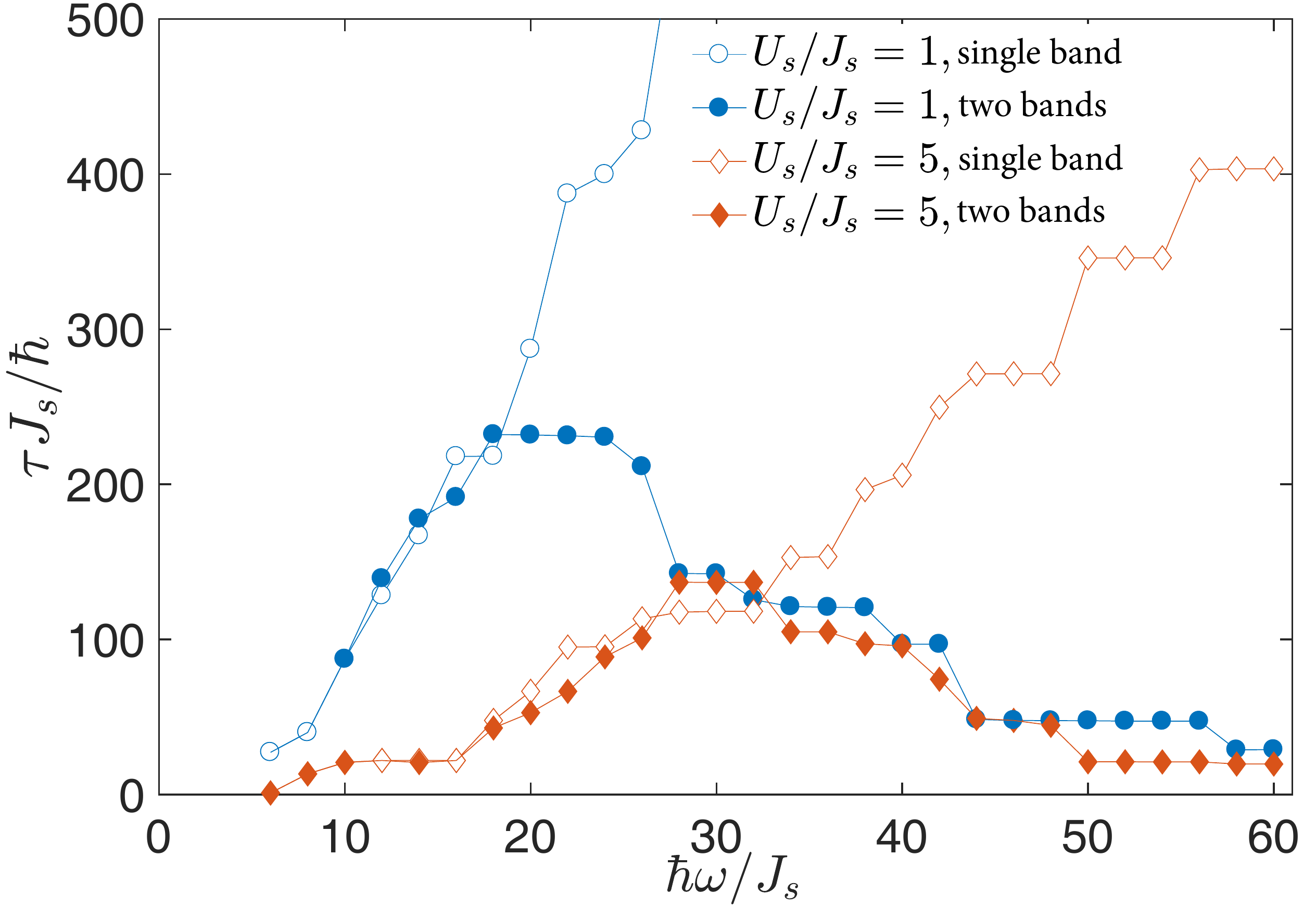}
\caption{Heating time versus driving frequency both for the 
single-band Hamiltonian (empty symbols) and the two-band Hamiltonian (full symbols)
for a system of $N=4$ particles on $M=8$ sites (corresponding to 16 single
particle-states) with $V_{0}/E_{R}=14$, $K/\hbar\omega=4$, and two different
interaction strengths.}
\label{fig:HeatingComp}
\end{figure}

In Fig.~\ref{fig:HeatingComp} we  plot the heating time $\tau$ versus the driving
frequency for a system of $N=4$ particles on $M=8$ sites (corresponding to 16
single-particle states) with lattice depth $V_0/J=14$. For strong driving,
$K/\hbar\omega=4$, and two different interaction strengths, $U_s/J_s=1$ and 
$U_s/J_s=5$, we compare the heating times obtained 
from the single-band model (\ref{eq:Hsb}) (open circles) to those obtained from
the two-band model (\ref{eq:tb}) (filled circles). As expected, we can observe 
that, while the coupling to the $p$ band does not influence the heating time for 
low frequencies, it becomes dominant for large driving frequencies. For the 
two-band model the interplay between intraband and interband heating gives rise 
to a maximum of the heating time, $\tau_\text{opt}$, at some optimal intermediate 
driving frequency $\omega_\text{opt}$. 
For the larger interaction strength $\tau_\text{opt}$ is lower and occurs at
a larger frequency. 

\begin{figure}[t]
\includegraphics[width=8.6cm]{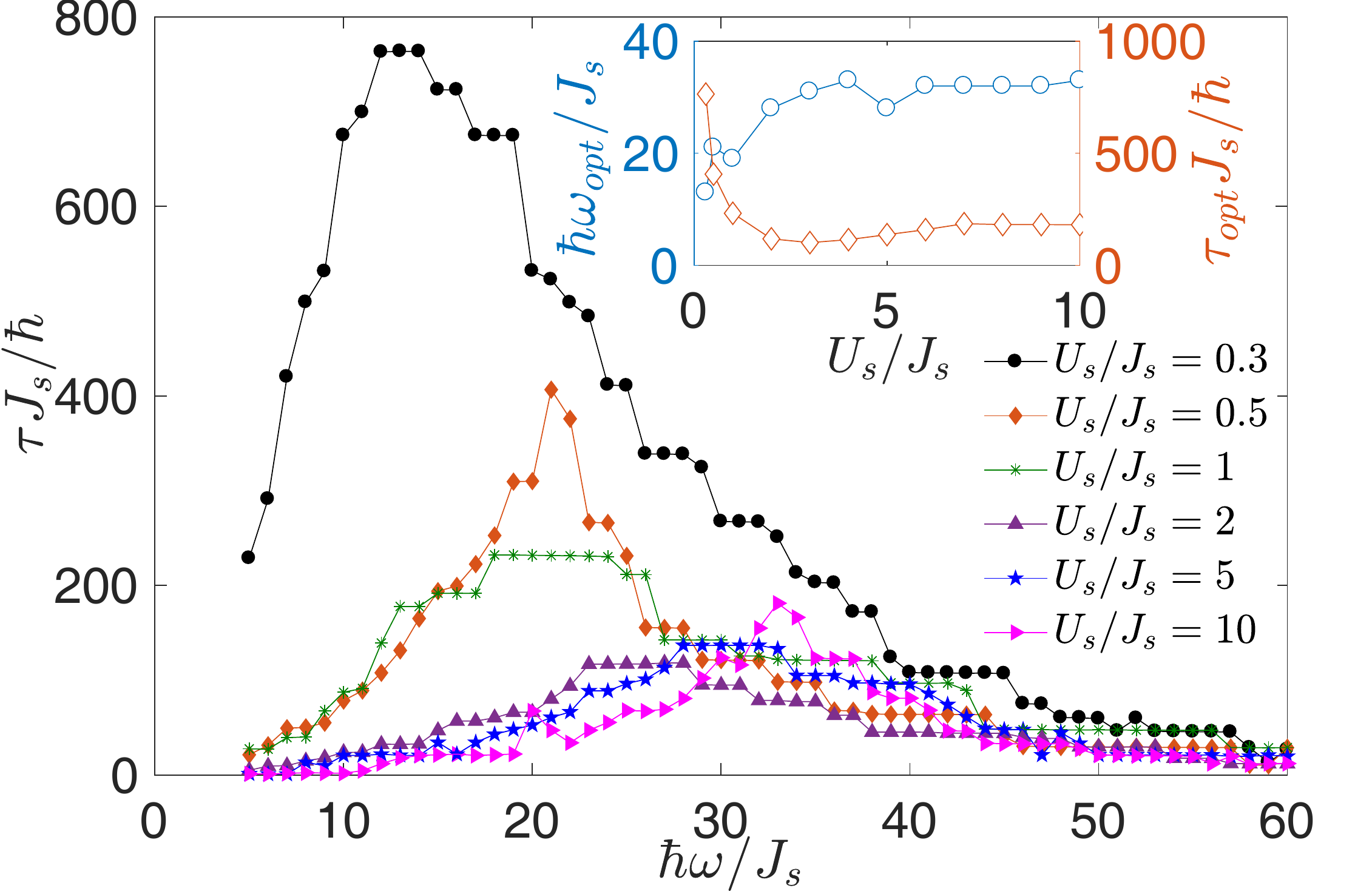}
\caption{Heating time $\tau J_s/\hbar$ versus driving frequency $\hbar\omega/J_s$ 
for the two-band model with different interaction parameters $U_s/J_s$, for $N=4$,
$M=8$, $V_{0}/E_{R}=14$, and $K/\hbar\omega=4$. The inset shows the optimal 
(maximum) heating time $\tau_\text{opt}$ (diamonds) and the corresponding optimal 
frequency $\omega_\text{opt}$ versus $U_s/J$.}
\label{fig:HeatingU}
\end{figure}

To study the impact of interactions in more detail, we compare the 
frequency-dependent heating times for various interaction strengths $U_s/J_s$ in 
Fig.~\ref{fig:HeatingU}. 
The inset shows the optimal (maximum) heating time $\tau_\text{opt}$ (diamonds, 
right axis) and the corresponding optimal driving frequency $\omega_\text{opt}$
(circles, left axis) versus $U_s/J_s$. We observe a significant reduction of
$\tau_\text{opt}$ combined with an upshift of $\omega_\text{opt}$, when 
increasing the interaction strength $U_s/J_s$ up to values of about 3. 
Both the shift of $\omega_\text{opt}$ and the noticeable reduction of
$\tau_\text{opt}$ for the single-band model (Fig.~\ref{fig:IntraOmega}) suggest
that increasing the interactions mainly enhances intraband heating, so that 
intraband heating becomes the dominant heating processes limiting $\tau$ up to 
larger values of $\omega$. For values of $U_s/J_s$ that are larger than 3 both 
$\tau_\text{opt}$ and $\omega_\text{opt}$ approximately saturate. 
We attribute this favorable behavior to the reaching of the strongly 
interacting regime $U_s/|J_s^\text{eff}|\approx 2.5 (U_s/J_s)\gg1$ in the lowest 
band. Here the kinetic energy of the particles is not sufficient anymore to 
induce changes in the site occupations that are associated with a change of 
interaction energy (for an initial state without multiply occupied sites this 
regime corresponds to the hard-core boson limit). Once this regime is reached, 
the physics within the lowest band does not change much anymore, when the 
interactions are increased further, which is consistent with to the observed 
saturation. This argument holds until eventually for even stronger 
interactions, $U_{sp}\sim \Delta$, deviations due to interband coupling will
make themselves felt.

\begin{figure}[t]
\includegraphics[width=8.6cm]{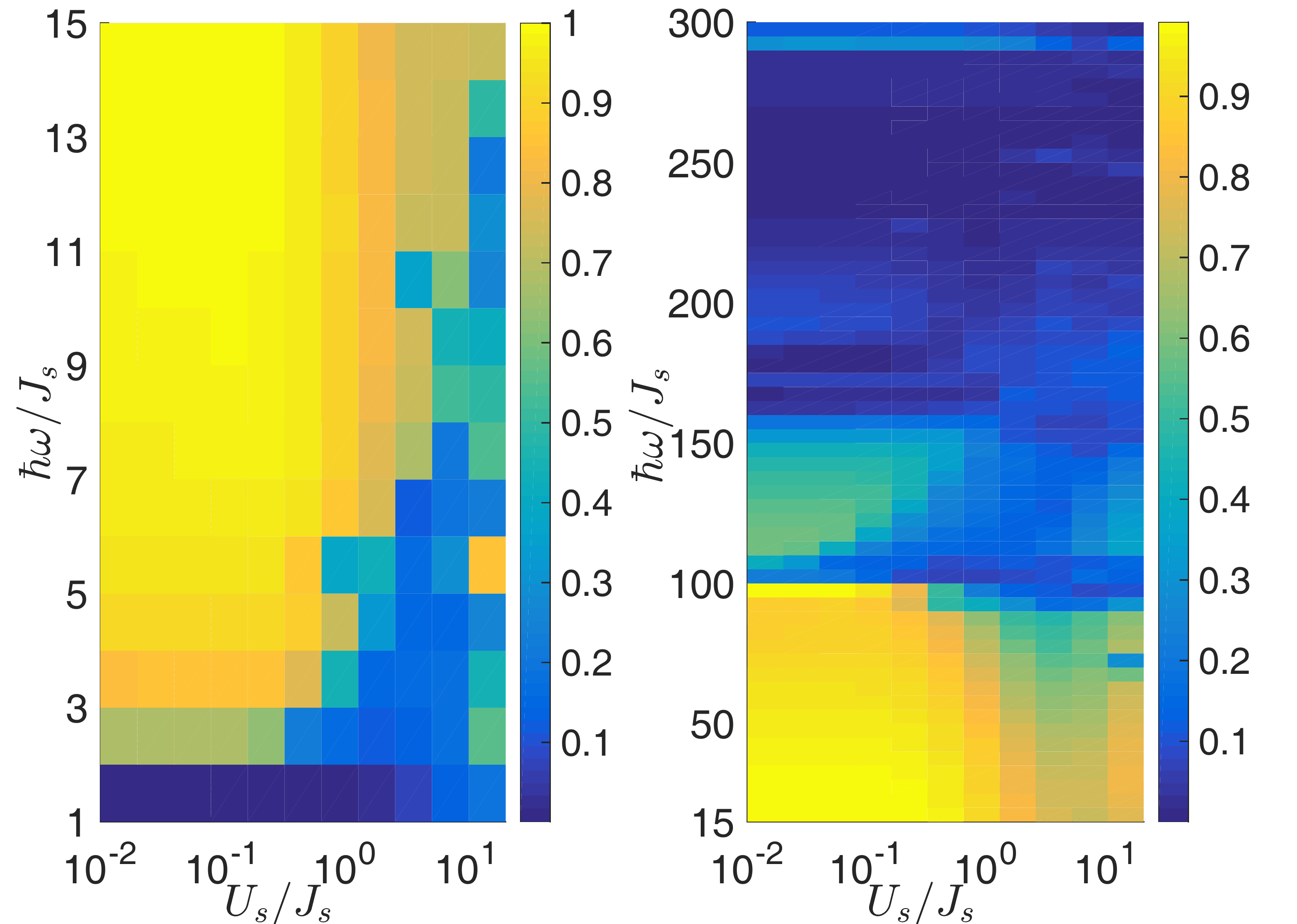}
\caption{Map of $n_0(t)/n_0(0)$ at time $t=100$ ms for the two-band model versus
driving frequency and interaction strengths, with $N=4$, $M=8$, $V_0/E_R=10$,
and $K/\hbar\omega = 1.5$. Here $n_0(t)$ is defined in Eq.~(\ref{eq:n0}).
We assumed a recoil energy of $E_R=3.33 2\pi\hbar$ kHz, a typical value for an 
experiment with $^87$Rb atoms, for which the chose time span corresponds to
$t J_s/\hbar\approx 40$ tunneling times. The driving strength corresponds to an 
effective tunneling matrix element of $J_s^\text{eff}\approx 0.5J_s$.}
\label{fig:Map}
\end{figure}

In Fig.~\ref{fig:Map} we depict the lowest-band zero-quasimomentum occupation
$n_0(t)$ in units of its initial value $n_0(0)$ at time 
$t\approx 40 \hbar/J_s$. This time is chosen to be large compared to the 
tunneling time $\hbar/J_s$, which is the relevant time scale for experiments.  
It is plotted versus the interaction strength $U_s/J_s$ and the driving frequency
$\hbar\omega/J_s$, where the low-frequency regime is shown in the left panel, 
while results for higher driving frequencies are given in the right panel. In 
the underlying simulations, we have considered a lattice depth of
$V_0/E_\text{R}=10$ and a driving strength of $K/\hbar\omega=1.5$, which 
is smaller than the one used previously and does not induce a sign change of
the effective tunneling matrix element (\ref{eq:Jeff}),
$J^\text{eff}_s\approx 0.51J_s$. 
The latter implies that on the the level of the effective Hamiltonian, the quench
induced when switching on the driving, does not correspond to an inversion of the
effective dispersion relation, but rather to a reduction of the band width by a
factor of one half. On the level of the time-averaged Hamiltonian (\ref{eq:Heff}),
this rather mild quench will excite the system only weakly, so that the 
occupation $n_0(t)/n_0(0)$ will retain a rather large value also 
during the dynamics following the quench. Thus, a significant reduction of
$n_0(t)/n_0(0)$ indicates unwanted driving-induced heating. Note also that (for 
fixed $K/\hbar\omega$) the ideal dynamics generated by $\Ho_\text{eff}$, and 
thus also $n_0(t)/n_0(0)$, should be independent of the driving frequency. 
Therefore, also any frequency dependence of $n_0(t)/n_0(0)$ must be viewed
as a deviation from the target dynamics generated by $\Ho_\text{eff}$. 

In Fig.~\ref{fig:Map}, we find signatures of heating in the form of a significant 
reduction of $n_0(t)/n_0(0)$ in various regimes. In the regime of weak 
interactions $U_s/J_s\ll1$, heating is visible both for too low frequencies, when
$\hbar\omega \sim J_s$, as well as for too high frequencies, when
$\hbar\omega\sim \Delta$ (with $\Delta/J_s\sim 250$ for the given lattice depth).
When the interband interactions $U_s$ become larger than the interband tunneling
$J_s$, low frequency heating sets in already for larger $\hbar\omega$, in 
accordance with condition (\ref{eq:lf}). At the same time, we can also observe
that interband heating at large frequencies is enhanced in the presence of
interactions. For the chosen lattice depth of $V_0/E_\text{R}=10$, we observe
that strong interactions $U_s/J_s\gg1$ lead to significant heating at any 
frequency.

\begin{figure}[t]
\includegraphics[width=8.6cm]{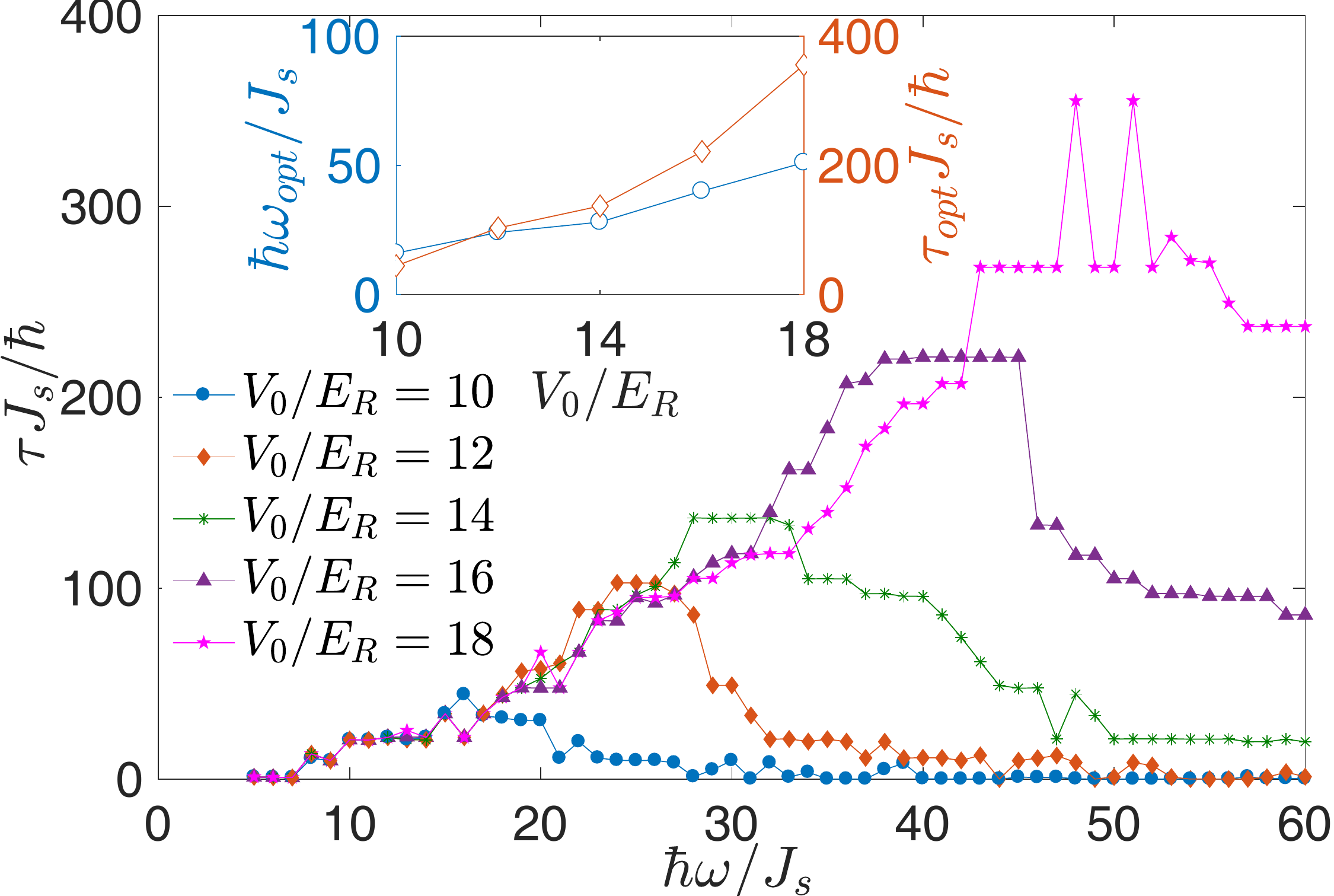}
\caption{Heating time $\tau J_s/\hbar$ versus driving frequency 
$\hbar\omega/J_s$ for the two-band model with 
different lattice depths $V_{0}/E_{R}$, for $N=4$, $M=8$, $U_s/J_s=5$, and
$K/\hbar\omega=4$. The inset shows the optimal 
(maximum) heating time $\tau_\text{opt}$ (diamonds) and the corresponding optimal 
frequency $\omega_\text{opt}$ versus $V_0/E_R$.}
\label{fig:HeatingV}
\end{figure}

The dependence of the heating time $\tau J_s/\hbar$ on the lattice depth
$V_0/E_\text{R}$ is investigated in detail in Fig.~\ref{fig:HeatingV}, where we 
plot the scaled heating time $\tau J_s/\hbar$ versus $\hbar\omega/J_s$ for 
various values of $V_0/J_s$ and for $U_s/J_s=5$ as well as $K/\hbar\omega=4$. 
The inset shows $\tau_\text{opt} J_s/\hbar$ and $\hbar\omega_\text{opt}/ J_s$ 
versus $V_0/E_\text{R}$. We can observe that both $\tau_\text{opt} J_s/\hbar$ 
and $\hbar\omega_\text{opt}/J_s$ increase with the lattice depth. The main figure 
shows that this behavior is associated with a significant reduction of heating 
for large $\hbar\omega/J_s$. Let us discuss this behavior in more detail. 

First, we can notice that the intraband dynamics, described by the single-band 
Hamiltonian (\ref{eq:Hsb}) and measured in the natural unit of the tunneling time
$\hbar/J_s$, is determined by the dimensionless ratios $U_s/J_s$,
$\hbar\omega/J_s$, and $K/\hbar\omega$, which we kept fixed in our simulations 
when increasing the lattice depth $V_0/E_\text{R}$. This choice of fixed parameters
is natural from the point of view of quantum simulation, where we wish to 
engineer the properties of the lowest band described by the approximate effective 
Hamiltonian (\ref{eq:Heff}). It explains why for small $\hbar\omega/J_s$, for
which interband coupling is negligible, the dimensionless heating time
$\tau J_s/\hbar$ is hardly influenced by the lattice depth. This can be seen from
the fact that all curves in Fig.~\ref{fig:HeatingV} agree up to the point 
($\sim \hbar\omega_\text{opt}/J_s$), where $\tau J_s/\hbar$ starts to be reduced
by interband processes.

We can, moreover, observe in Fig.~\ref{fig:HeatingV} that the interband heating, 
which is responsible for the reduction of $\tau J_s/\hbar$ at large frequencies, 
is significantly reduced with increasing lattice depth. This behavior results from
the interplay of various effects. On the one hand, with increasing lattice depth 
$V_0/E_\text{R}$ the band separation $\Delta/E_\text{R}$ increases, whereas the 
interband coupling parameter $\eta$ decreases (Fig.~\ref{fig:parameters}). Both
effects tend to reduce interband heating. An additional and much stronger 
reduction of interband heating will, however, results from the exponential 
suppression of the tunneling parameter $J_s$ with the square root of the 
lattice depth $V_0/E_\text{R}$ (Fig.~\ref{fig:parameters}). Namely, since we 
keep the dimensionless ratio $\hbar\omega/J_s$ fixed (taking the point of view of 
quantum simulation, as explained in the previous paragraph), the number
$n_\text{ph}$ of photons (i.e.\ energy quanta $\hbar\omega$) needed to overcome 
the band separation $\Delta$, $n_\text{ph}\approx\Delta/(\hbar\omega)$, will 
strongly increase with the lattice depth. This, in turn, implies a very strong 
suppression of interband heating, since we expect an exponential suppression of 
interband transitions with $n_\text{ph}$ \cite{StraeterEckardt16,WeinbergEtAl15}. 

The effects described in the previous paragraph explain a strong increase of 
the heating time $\tau$ with increasing lattice depth. However, from the point 
of view of quantum simulation, we have to compare the heating time to the relevant
experimental time scale, given by the tunneling time $\hbar/J_s$. This is why 
here we are always plotting the scaled heating time $\tau J_s/\hbar$. Therefore,
when increasing the lattice depth, the expected strong increase of $\tau$ 
directly competes with the exponential increase of $\hbar/J_s$ with the square 
root of the lattice depth. The results presented in Fig.~\ref{fig:HeatingV} 
clearly show that the former effect wins over the latter one, so that, all in 
all, $\tau J_s/\hbar$ is reduced when the lattice depth $V_0/E_\text{R}$ is 
raised. We can, thus, see a noticeable increase of the optimal heating time 
$\tau_\text{opt}J_s/\hbar$ (shown in the inset of Fig.~\ref{fig:HeatingV}) with
$V_0/E_R$. 

While the results of Fig.~\ref{fig:HeatingV} imply that driving-induced heating 
can effectively be reduced by raising the lattice depth, this possibility is 
limited by non-driving-induced heating processes, originating, e.g., from 
three-body collisions, scattering with background particles, or noise. Namely,
the tunneling time, which increases with the lattice depth, has to remain short
compared to the the time scale $\tau_0$ associated with such background heating. 
In turn, this means that by increasing $\tau_0$ by reducing non-driving induced 
heating, the experimentalist can also reduce driving-induced heating. This is 
a major result of this article.

\begin{figure}[t]
\includegraphics[width=8.6cm]{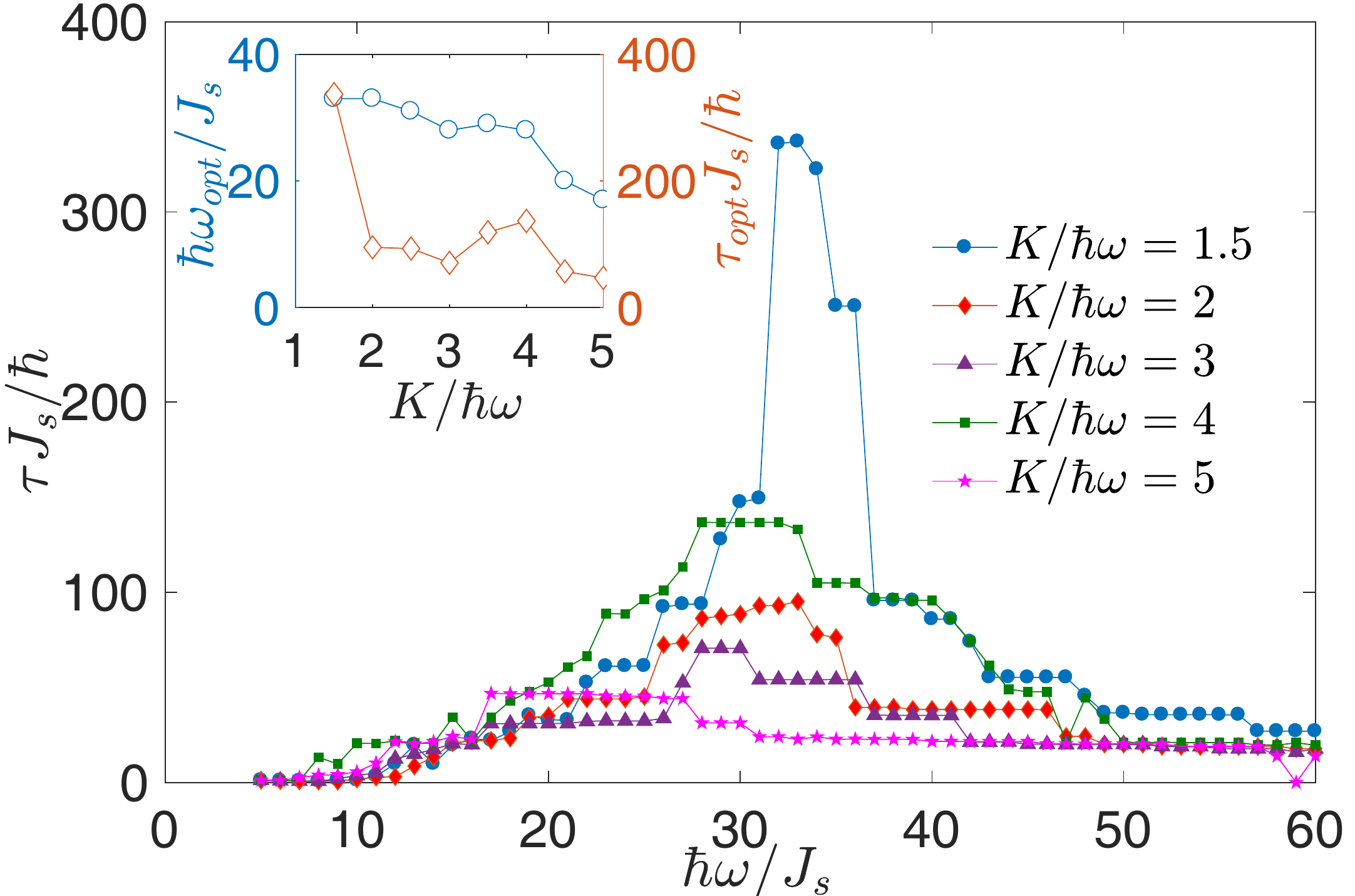}
\caption{Heating time $\tau J_s/\hbar$ versus driving frequency $\hbar\omega/J_s$for the two-band model with different driving amplitudes $K/\hbar\omega$, for
$N=4$, $M=8$, $V_{0}/E_{R}=14$,
and $U_s/J_s=5$. The inset shows the optimal 
(maximum) heating time $\tau_\text{opt}$ (diamonds) and the corresponding optimal 
frequency $\omega_\text{opt}$ versus $K/\hbar\omega$.}
\label{fig:HeatingK}
\end{figure}

Let us, finally, also have a look at the dependence of the heating time on the
driving strength. In Fig.~\ref{fig:HeatingK} we plot $\tau J_s/\hbar$ versus
$K/\hbar\omega$ for a system with $V_0/E_\text{R}=14$ and $U_s/J_s=5$. We focus
on values of $K/\hbar\omega$ that are interesting for Floquet engineering (i.e.\ 
that are large enough to achieve a significant modification of $J^\text{eff}_s$ 
and not much larger than required for tuning $J^\text{eff}_s$ to negative 
values). For the smallest considered driving strength of $K/\hbar\omega=1$ a 
narrow window of frequencies is found for which the heating time takes large 
values of more than 300 tunneling times. This window disappears for stronger 
driving. Note that we do not find a simple monotonous decrease of the heating 
time with respect to the driving strength. We attribute this observation to the 
non-monotonous behavior of the finite-frequency components
$\propto e^{i m \omega t}$ of the time-dependent Hamiltonian (\ref{eq:Hprime}) 
in the rotating frame [as well as of the corresponding two-band Hamiltonian]. 
Namely, these terms, which describe heating processes beyond the rotating wave 
approximation (\ref{eq:Heff}) where the system exchanges $m$ energy quanta 
$\hbar\omega$, involve Bessel-function expressions
$\mathcal{J}_m(K/\hbar\omega)$ that depend in a non-monotonous way on the 
driving strength $K/\hbar\omega$.

\section{Conclusions\label{sec:conclusions}}

In summary, we have investigated the conditions for Floquet engineering in 
optical lattices. In particular we were interested in the existence of a 
frequency window where both low-frequency intraband heating and high-frequency 
interband heating is suppressed on a time scale $\tau$ that is large compared 
to the tunneling time. Considering the concrete example of a small 
one-dimensional system of interacting bosons in a shaken optical lattice, we 
presented numerical results that show that such a frequency window exists for 
sufficiently deep lattices. The maximum ratio of heating and tunneling time, 
$\tau_\text{opt} J_s/\hbar$, (which is found for an optimal intermediate 
driving frequency $\omega_\text{opt}$) is found to increase with the lattice 
depth. This result, which is not obvious since also the tunneling time 
increases exponentially with the lattice depth, implies that we can reduce 
driving-induced heating, by simply ramping up the lattice depth. However, we 
have pointed out that this strategy is limited to lattice depths for which the 
tunneling time is still much smaller than the time scale $\tau_0$ for 
non-driving-induced background heating. Thus, the larger the time scale for 
such background heating, the more we can reduce also driving induced heating. 

We have also found that ramping up the interaction strengths, driving-induced
heating is significantly enhanced, until a saturation value is reached roughly 
when the ratio $U_s/J_s$ reaches values of 3. This saturation-behavior is a
promising result regarding the possibility of Floquet engineering of strongly
correlated states of matter such as fractional Chern insulators
\cite{GrushinEtAl14, AnisimovasEtAl15, RaciunasEtAl16, RaciunasEtAl18}.

An interesting direction for future work concerns the role of disorder. It has
been argued that many-body localization can protect the driven system against
unwanted heating associated with deviations from the high-frequency approximation
\cite{LazaridesEtAl15, PonteEtAl15}. Roughly speaking, within the localization
length, the system is not able to create excitations of a sufficiently large
energy $\hbar\omega$. The mechanism is crucial also for the stabilization of 
discrete time crystals \cite{KhemaniEtAl16,ElseEtAl16,ChoiEtAl17, ZhangEtAl17, 
Sacha_Zakrzewski2017, KhemaniEtAl19, ElseEtAl19}.
However, disorder-induced localization cannot be expected to prevent the system
also against unwanted heating associated with deviations from the low-frequency 
approximation. Unwanted resonant multi-photon excitations to states above the gap
can still occur. It is an interesting question, in how far the corresponding
heating rates are influenced by disorder-induced localization.

\begin{acknowledgments}
We thank Christoph Str\"ater for providing the band structure data. This work 
was supported by the German Research Foundation DFG via the Research Unit 
FOR2414 (under Project No. 277974659).
\end{acknowledgments}

\bibliography{mybib}

\end{document}